# Fast and Sensitive Terahertz Detection in a Current-Driven Epitaxial-Graphene Asymmetric Dual-Grating-Gate FET Structure

Koichi Tamura*[1], Chao Tang[1], Daichi Ogiura[1], Kento Suwa[1], Hirokazu Fukidome[1],

Yuma Takida[2], Hiroaki Minamide[2], Tetsuya Suemitsu[3], Taiichi Otsuji**[1], and Akira

Satou[1]

[1] *Research Institute of Electrical Communication, Tohoku University, Sendai 980-8577,*

*Japan*

[2] *RIKEN Center for Advanced Photonics, RIKEN, Sendai, Miyagi, 980-0845, Japan*

[3] *New Industry Creation Hatchery Center, Tohoku University, Sendai 980-0845, Japan*

**Abstract**: We designed and fabricated an epitaxial-graphene-channel field-effect transistor (EG-FET) featured by the asymmetric dual-grating-gate (ADGG) structure working for a current-driven terahertz detector, and experimentally demonstrated a 10-ps order fast response time and a high responsivity of 0.3 mA/W to the 0.95-THz radiation incidence at room temperatures. The ADGG- and the drain-source-bias dependencies of the measured photoresponse showed a clear transition between plasmonic detection under periodic electron density modulation conditions with depleted regions and photothermoelectric detection under highly doped conditions without depleted regions. We identified the photothermoelectric detection that we observed as a new type of unipolar mechanism in which only electrons or holes contribute to rectifying the THz radiation under current-driven conditions. These two detection mechanisms coexist in a certain wide transcendent range of the applied bias voltages. It was also clearly manifested that the temporal photoresponse of the plasmonic and photothermoelectric detection are comparably fast on the order of 10 ps, whereas the maximal photoresponsivity of the photothermoelectric detection is almost twice as high as that of the plasmonic detection under the applied biases conditions. These results suggest that the ADGG-EG-FET THz detector will be promising for use in 6G- and 7G-class high-speed wireless communication systems.

\*  koichi.tamura.t8@dc.tohoku.ac.jp

\*\* Corresponding Author: otsuji@riec.tohoku.ac.jp

Research Institute of Electrical Communication, Tohoku University. 2-1-1 Katahira, Aobaku, Sendai, 980-8577 Japan

**Introduction**: Terahertz (THz) electromagnetic waves have great potentials to be utilized for applications of large-capacity, ultrahigh-speed wireless communication technologies in 6G toward 7G [1,2]. To realize such high-speed communication systems, highly sensitive, fast response, room-temperature detectors operating in the THz and sub-THz ranges are the key elements [3,4]. However, there are many performance limitations in the currently available THz detectors [5,6]. Two-dimensional (2D) plasmons have attracted increasing attention as a promising mechanism of highly sensitive, fast response THz detection [6-16]. Particularly graphene Dirac plasmon (GDP) [17-19] is believed to be one of the most promising physical principles for breaking through the technological limit on room-temperature, fast, sensitive THz detection capable for the 100-Gbit/s class high-data-rate coding of THz- and sub-THz radiation incidence in the next-generation 6G- and 7G-class wireless communications systems [20]. Graphene has also served fast photothermoelectric THz detection [21-25] thanks to its superior carrier transport and phononic properties [26]. In this paper, we design and fabricate an epitaxial-graphene-channel field-effect transistor (EG-FET) featured by the authors' original asymmetric dual-grating-gate (ADGG) structure [27] working for a current-driven terahertz detector with applied non-zero drain-source bias voltages, and experimentally demonstrate a 10-ps-order fast temporal response and a high responsivity of 0.3 mA/W (equivalently 12 mV/W under the 50-$\Omega$-loaded condition and 84 mV/W under the high (~1 M$\Omega$) loaded impedance condition) to the 0.95-THz radiation incidence at room temperatures. The ADGG- and the drain-source-bias dependencies of the measured photoresponse show a clear transition between plasmonic detection and photothermoelectric detection while preserving the fast response speed. The experiments also benchmark the temporal photoresponse of the plasmonic and photothermoelectric detection to be comparably fast on the order of 10 ps.

**Experimental Methods**: The bird's-eye view of the ADGG-EG-FET structure is schematically shown in Fig. 1. The channel consists of a-few-layers' epitaxial graphene that was thermally decomposed from a C-face SiC substrate [28–32]. The gate stack was formed with a 40-nm-thick SiN dielectric layer deposited on the graphene channel layer using plasma-enhanced chemical vapor deposition (PE-CVD) [32,33]. The gate metal electrode was formed in the ADGG structure by using electron-beam lithography, an electron-beam evaporator, and a standard lift-off process [32,33]. The scanning electron microscopy (SEM) image of a fabricated ADGG-EG-FET is depicted in Fig. 2(a). The ADGG electrodes consist of two interdigitated grating-shaped metals with different grating finger widths of 500 nm ($L_{g1}$) and 800 nm ($L_{g2}$), were laid out with asymmetric distances of 500 nm ($d_1$) and 800 nm ($d_2$) to the left-side and right-side adjacent fingers, respectively. The source and drain electrodes were formed on top of the graphene channel in planar [34] and edge [35] ohmic contacts. The crystallinity of the graphene layer was confirmed by Raman spectroscopy shown in Fig. 2(b). The G and G' bands at 1,590 cm$^-$$^1$ and 2,700 cm$^{-1}$ were clearly identified in the Raman spectra; oppositely, the defect-oriented D band at 1,350 cm$^{-1}$ was as weak as the background noise floor, indicating the favorable high-quality crystallinity of the graphene layer [36]. The ratio between the intensity of G and G' peaks was about 3.5, indicating the samples to be a-few-layers graphene within three layers [37]. The surface morphology of the graphene layer was characterized by atomic force microscopy (AFM) shown in Fig. 2(c). Crystal domains with a size of several micrometers were identified in the AFM image [38].

To investigate the hysteresis in electrical properties of the as-fabricated EG-FET, we



conducted the electrical direct current (DC) measurement using a semiconductor parametric analyzer. The DC drain-source current versus the gate bias scanning from 0 V to -20 V back and forth was measured under the condition of $V_{g2} = 0$ V and $V_{ds} = 0.1$ V. As shown in Fig. 3, The ambipolar characteristics near the Dirac voltage ($V_{Dirac}$, defined as the charge neutrality voltage point) in negative gate voltage was observed. The difference of current between the forward and backward bias applying was merely several µAs. Although the experiments under the condition of the p-type high-current operation region with the negative gate voltages lager than $V_{Dirac}$ were not able to be conducted due to the gate breakdown limitation, such a wide shift of the $V_{Dirac}$ is considered to be due to relatively high unintentional n-type doping to the graphene channel that occurred during the SiN insulator deposition process using the PE-CVD [32]; on the other hand, there are almost no hysteresis experimentally observed in the situation of the high drain source bias which will be discussed in the paragraph of the mechanism of photothermoelectric detection.

When THz radiations are incident on the surface of the device, the ADGG electrodes work as a broadband antenna that can efficiently convert the incident THz photons to the GDPs [27,39]. When one grating gate of the ADGG electrodes G1 is electrically biased at a high voltage whereas another gate G2 is biased low at the Dirac voltage to deplete the carriers, the channel underneath the high-biased G1 gate finger becomes a plasmonic cavity, working as a plasmonic detector producing a rectified dc photocurrent due to the hydrodynamic nonlinearity of the GDPs [15,16,27]. The depleted channel underneath the low-biased G2 gate finger becomes a highly resistive load, working as a transducer to produce a photovoltage from the photocurrent [11,16]. Due to the periodic arrangement of such a unit pair of the plasmonic cavity and the resistive load in the 'ADGG' structure, the



photovoltage generated in each unit pair is summed up in a cascading manner, resulting in a highly sensitive THz detection. It is noted that the difference between $d_1$ (= 500 nm) and $d_2$ (= 800 nm) or the asymmetricity ($d_1/d_2 \neq 1$) is the key to unbalancing the boundary conditions at the left side and right side of the plasmonic cavity so that the plasmonic displacement current flows from source/drain to drain/source become unbalanced, resulting in rectified DC photovoltaic output at the drain terminal. The length of the high-biased gate ($L_{g1}$ = 500 nm in the above context) determines the plasmonic resonant mode frequency, whereas the length of the low-biased gate ($L_{g2}$ = 800 nm in the above context) determines the load resistance value.

We measured the temporal response of the photovoltage output from the drain electrode in response to the pulsed quasi continuous-wave (CW) radiation incidence centered at 0.95 THz at room temperature. The THz detection is conducted and implemented with an injection-seeded THz parametric generator (is-TPG) [40] utilized as the THz radiation incident source (Fig. 4). The is-TPG generated pulsed-CW THz radiation with an envelope pulse width of 155 ps and a repetition of 200 Hz [41]. The envelope pulse width of 155 ps was identified by using an optically up-converted cross-correlation method with a sub-ns Nd:YAG infrared pump pulse whose wavelength was centered at 1,064 nm as described in Ref. [41]. The THz waves that were output from the is-TPG traveled in free space, were focused by a Tsurupica lens having a focal distance of 100 mm, and were directed via an ITO mirror to the sample surface placed at the focal point. The radiation incidence energy was 137 nJ/envelope (peak power of ~911 W). A set of the RF probes were contacted to the ADGG-EG-FET electrode pads to apply the bias voltages (the drain-to-source bias $V_{ds}$ and two ADGG biases $V_{g1}$ and $V_{g2}$, respectively). To observe the temporal photoresponse waveform without distortions



caused by the multireflection between the device output and the far end of the measurement equipment, we used a 50-Ω-impedance measurement setup consisting of a 50-Ω-input-impedance, a 22-dB gain wideband preamplifier, a 50-Ω-input-impedance, 33-GHz bandwidth digitizing oscilloscope, and a 1-m-long, 50-Ω-coaxial-cabled transmission line to connect the device output terminal and the oscilloscope. Compared to a high-impedance measurement setup that is frequently utilized in static DC-voltage photoresponse measurement, the measured photovoltage under the 50-Ω-loaded condition becomes small by a factor of the voltage divider ratio between the internal channel resistance $R_{ch}$ (~300 Ω) and the load resistance $R_L$ (= 50 Ω) given by $R_L/(R_{ch}+R_L)$ ~ 0.14 in this experiment.

**Results:** Firstly, we conducted experiments at a sufficiently doped voltage of +15 V for $V_{g1}$ and at the Dirac voltage (= $V_{\text{Dirac}}$, the charge neutrality point) for $V_{g2}$ to make sure the channel regions underneath G1 (G2) were sufficiently doped (depleted), where the device is only responsible for plasmonic rectification of THz radiation theoretically. Besides, the $V_{ds}$ was biased at 1 V to drive the current in the channel. Under the condition $V_{g2} = -15.5$ V in Fig. 5, we confirmed a clear photovoltaic response at room temperature, indicating the ADGG-EG-FET works properly as a plasmonic THz detector. The temporal photoresponse showed a tail-free high-fidelity waveform with an FWHM (full width at half maximum) value of 199 ps traceable to the incident pulsed quasi-CW radiation envelope whose FWHM value was characterized to be 155 ps by using an optical up-conversion-based cross-correlation method as described in Ref. [41]. The observed pulse width of 199 ps in FWHM was a bit wider than the envelope width of the is-TPG radiation incidence that was characterized to be 155 ps by using a different optical nonlinear cross-



correlation method [41]. The discrepancy between them might include several systematic factors caused by the different routes of pulse-width characterization. As in the photoresponse of the ADGG-EG-FET, the energy relaxation processes of hot carriers in graphene activated by highly intense THz radiation incidence might be another perturbative factor, which will be a future study. Nevertheless, to the best of the authors' knowledge, this is the first experimental demonstration of 10-ps-order fast temporal photoresponse of graphene-based THz detectors at room temperatures.

Secondly, after the experiments under the plasmonic rectification condition ($V_{g2} = V_{\text{Dirac}}$), we measured the photoresponse with increasing $V_{g2}$ from $-15.5$ V (= $V_{\text{Dirac}}$) to $+15$ V (= the $V_{g1}$) under the fixed $V_{g1}$ (= $+15.5$ V) and $V_{ds}$ (= $+1.0$ V) condition. As shown in Fig. 5, the measured temporal photovoltage output increased with increasing $V_{g2}$ and corresponding electron densities underneath G2 while preserving the high-speed response. The peak values of the photovoltage versus $V_{g2}$ are plotted in Fig. 6(a). With increasing $V_{g2}$, the photoresponse increased to saturate at ~ -5 V. We claim that the observed increase in the photo-responsivity is due to a new type of unipolar photothermoelectric effect assisted by electrostatic carrier drift/diffusion, which will be discussed in the **Discussion section**.

The gate biases G1 and G2 conditions of plasmonic/photothermoelectric rectification are shown in Fig.6(b) and 6(c). To confirm the behavior of photoelectrons due to the photothermoelectric process, we fixed $V_{g1} = V_{g2} = 0$ V, which ensured the graphene channel area was entirely doped sufficiently, then increased $V_{ds}$ from 0 to $+1.5$ V. The temporal photoresponse waveforms under typical non-zero $V_{ds}$ bias voltages conditions were measured as plotted in Fig. 7(a). The temporal photoresponse preserved its waveform with an FWHM value of 199 ps independent of the applied non-zero $V_{ds}$



bias voltages.

The peak values of the temporal photoresponse under the 50-$\Omega$-loaded condition shown in Fig. 7(a) are plotted in Fig. 7(b) as a function of $V_{ds}$. The output photovoltage increased linearly with increasing $V_{ds}$. More importantly, when $V_{ds} = 0$ V preventing from the photothermoelectric rectification operation, no photoresponse was observed. This is clear evidence that the photoresponse observed in Figs. 7(a) and 7(b) under the fully doped conditions were attributed to the photothermoelectric rectification effect.

**Discussion**: Besides the mechanism of THz detection by the GDPs, the experimental results suggest that the ADGG-EG-FET is also able to work as a current-driven photothermoelectric THz detector [21-25] thanks to the photo-Seebeck effect. When $V_{g2}$ is biased near the Dirac point, the carrier in the channel is not sufficient to drive the photothermoelectric effect; the plasmon rectification predominates the detection, which is shown in the blue part of Fig. 6(a). With increasing $V_{g2}$, the injected electron density increases sufficiently to create a certain amount of current driven by the potential slope from source to drain electrode, which is shown in the orange part in Fig. 6(a). In detail, when the drain terminal is DC-biased, the electric potential gets a slope along the channel, resulting in an asymmetric carrier diffusion under THz radiation incidence due to the photo-Seebeck effect along the channel with the help of the field-induced electrostatic drift/diffusion; the hot carriers photoexcited by THz electromagnetic radiation will diffuse to be biased in the direction of the potential slope. When all gate electrodes are zero-biased, the thermo-diffusion occurs isotopically and there is no specific fraction of the diffusion direction either to the source or to the drain; however, photothermoelectric detection makes it possible to conduct a "zero bias detection" thanks to the electrostatic



movement of the carriers due to the electrical potential slope along the channel. The electrostatic drift diffusion with a potential gradient makes asymmetrical carrier flow. In this regard, it might be said that this is an electrostatic-drift/diffusion-assisted photothermoelectric detection. Suppose the THz radiation spot size could be minimized far below the diffraction limit like a micrometer and swept from near the source to near the drain. In that case, the difference in the distance between the spot point and the drain/source terminal may produce a photovoltaic response (due to the difference of the thermal/electrical resistance between the two distant regions). If so, one can observe the PTE detection response even under the zero-drain bias condition. But our case is not such, but the spot covers the entire channel region so that the thermo-diffusion becomes totally isotropic and does not have any directional-dependent fractions, which should be evidence that our experiment (photoresponse only under non-zero drain bias conditions) can be interpreted as a kind of "photothermoelectric" detection.

As for the "zero bias detection" using the plasmonic mechanism, the Dirac voltage is deeply shifted to the negative region, and the spatial carrier density distribution along the channel under positively biased gate conditions is rather monotonic and not so periodically modulated as the photoresponse as a function of $V_{DS}$ gets saturated at $V_{G2} \sim$ 0 V and beyond. Thus, it is hard to get a zero-bias detection as plasmonic detection makes it possible. As a consequence, a photovoltaic signal is an output from the drain terminal in response to the THz radiation incidence. This is not similar to the standard photothermoelectric rectification in bipolar p-n junction diode structures[21-25] with different work function metals of the anode and cathode electrodes in which both electrons and holes contribute to the rectification function. Our current-driven ADGG-EG-FET contributes only unipolar carriers of hot electrons or hot holes excited by the



THz radiation incidence. Therefore, the observed THz radiation rectification mechanism is regarded as a new type of unipolar photothermoelectric detection driven by thermo-diffusive and field-induced electrostatic drift/diffusion to rectify the THz radiation. These results shown in Fig. 6 suggest that the plasmonic rectification and/or the photothermoelectric rectification take/takes place in the ADGG-EG-FET under THz radiation incidence depending on the ADGG biases conditions while preserving the 10-ps-order fast response speed performance and that these two effects coexist under a wide range of $V_{g1}$ and $V_{g2}$ conditions.

It is also worth noting that the plasmonic THz detection works well under the zero-drain biased condition with zero-power consumption [13-16] whereas the new type of unipolar photothermoelectric detection works only under non-zero drain biased condition with non-zero power consumption as demonstrated in Fig. 6(b). In this regard, the ADGG-EG-FET can work in both cases with and without any power supply via current-driven detection and zero-biased plasmonic detection.

The estimated intrinsic current responsivity of the ADGG-EG-FET detector, defined as the ratio of the photocurrent to the power of the incident THz wave to the active detector area, was ~0.3 mA/W. Correspondingly, the intrinsic voltage responsivity under a high loaded impedance condition, which is given by the product of the current responsivity $R_I$ and the channel resistance $R_{ch}$, was characterized to be 84 mV/W. To further investigate the level of responsivity in comparison with the recently published result with the maximal responsivity of 1.9 mA/W in an ADGG-EG-FET detector using the highest quality of exfoliated hBN/graphene/hBN van der Waals heterostructures at 0.3 THz at 300 K [16], the observed responsivity of 0.3 mA/W at 0.95 THz at 300 K is about 1/6 as high as that in the reported value at 0.3 THz at 300 K. The first factor that we



should consider is the difference in detection frequency. As is reported experimentally [11] and theoretically [42], the plasmonic ratchet effect and drag effect take different frequency dependence on responsivity, and the overall trend shows a monotonic decrease with increasing frequency. In the case of InGaAs/InAlAs/InP ADGG-HEMT detectors [11], the responsivity at 1 THz weakens by one order of magnitude from that observed at 0.3 THz. Our result obtained in ADGG-EG-FETs showed less attenuation of ~1/6, which is thought to be due to the superior transport properties of graphene Dirac fermions.

The second factor is considered to be the less crystallinity of epitaxial graphene damaged throughout a standard semiconductor integrated device processes including the PE-CVD gate stack process used in this work than that for the highest quality exfoliated graphene with less damaged exfoliation/transfer device processes reported in Refs. [15 and 16]. According to the previously published results [43,44], the relaxation time ($\tau$) of electrons can be estimated by the linear relationship between $\tau$ and the intensity ratio of Raman G and D peaks, which was calculated in the range of 8.25 to 25.30 for our epitaxial graphene, shown in Fig. 2(b), resulting in $\tau$ values in the range of 1 to 2.5 ps. The corresponding carrier mobilities were identified to be ~16,900~43,000 cm$^2$/Vs (the derivation is given in Supporting Information). Our graphene sample was fabricated using the thermal decomposition method from the $(000\bar{1})$ surface of a chemical and mechanical polished SiC wafer. This method is able to create high-quality graphene having carrier mobility of 100,000 cm$^2$/Vs even at room temperature, which is experimentally manifested by using Time- and Angle-resolved photoemission spectroscopy (T-ARPES) [45]. In this reported experiment, the *in-situ* measured nonequilibrium carrier energy relaxation delay time has been properly fitted using the parameters including the Fermi velocity and carrier mobility [45]. Moreover, in another EG-FET device fabricated on this



kind of graphene membrane in our previously published literature [46], the mobility is derived from the electric properties of EG-FET as 50,600–63,300 $cm^2/Vs$. According to these, our estimated value of mobility in this paper is reasonable. Compared with the mobility of ~38,000 $cm^2$/Vs in the ADGG-graphene-channel FET [16] fabricated using mechanical exfoliation and transfer process with an encapsulation sandwiched by two h-BN thin layers, the carrier mobility in the epitaxial graphene of this work is fairly comparable. From this result, such a process-dependent degradation of the quality factor will be minor. The further quantitative investigation will be given after future experimental works.

It was proved that the photothermoelectric effect could enhance the detection sensitivity in the current-driven ADGG-EG-GFET detector from the level of the plasmonic rectification effect without any deterioration of the fast response speed. The obtained 10-ps order fast detector response meets the requirements for the level of the fast, sensitive receiver to be applicable to 6G/7G-class next-generation THz wireless communication systems. Besides, all of the processes in the device fabrication are suitable for the mass-production implemented with present semiconductor integrated device processing technologies.

Finally, we estimated the noise equivalent power (*NEP*) taking account of both the thermal noise and the shot noise as follows [11]:

$$NEP = \frac{(N_{ThV} + N_{ShV})}{R_i}, \tag{1}$$

$$N_{ThV} = \sqrt{4k_B T_e / R_{ch}}, \tag{2}$$

$$N_{ShV} = \sqrt{2eI_{ch}}, \tag{3}$$

where $N_{ThV}$ and $N_{ShV}$ are the thermal noise and shot noise factors in the unit of A/$\sqrt{\text{Hz}}$, respectively, $k_B$ is the Boltzmann constant, $T_e$ is the electron temperature, $e$ is the elementary charge, $I_{ch}$ is the channel current, and $R_{ch}$ is the channel resistance. In these experiments, the output photovoltaic response was smaller than the true signal by a factor of the voltage divider ratio of $R_L/(R_{ch} + R_L)$ where $R_L$ is the load resistance of 50 $\Omega$. It is worth noting that even for use in case of the high-speed wireless transmissions wherein the highest signal integrity with the least distortion is a compulsory condition the high-impedance-loaded condition can work in a good sense if the interconnection length between the detector output and the input of the next-stage circuitry like a preamplifier could be managed within 100 micrometers (below 0.1 of the electrical wavelength under consideration) that is routinely obtainable in monolithic integrated circuit technology. Then, therefore, the photovoltages of the true signals under the high output impedance are ~7 times higher than the observed outputs.

The *NEP* and the intrinsic current responsivity $R_i$ versus $V_{ds}$ are plotted in Fig. 8. According to Eqs. (1-3) we obtained the minimum value of the *NEP* of 166 nW/$\sqrt{\text{Hz}}$ at the maximum $R_i$ ($R_v$) of 0.3 mA/W (84 mV/W) at room temperature. These responsivity and noise performances are comparably high among ever reported graphene THz detectors but are lower than those in InGaAs/InAlAs/InP ADGG-HEMTs. This is because the gapless graphene prevents complete depletion of carriers giving rise to relatively low channel resistance reflecting the low responsivity. The introduction of bilayer graphene to open the bandgap with an application of the vertical electric field intensity will improve the responsivity drastically to a certain extent without severe degradation of the speed performance.

**Conclusion**: We designed and fabricated our original ADGG-EG-FET structure working for a current-driven terahertz detector, and experimentally demonstrated a 10-ps-order fast temporal response, a high responsivity of 0.3 mA/W, and a *NEP* of 166 nW/$\sqrt{\,}$Hz over a wide variation of the gate and drain bias voltages at room temperatures. The ADGG-bias dependence of the measured photoresponse showed a clear transition between plasmonic and photothermoelectric detections while preserving the fast response speed. We identified the photothermoelectric detection that we observed as a new type of unipolar mechanism in which only electrons or holes contribute to rectifying the THz radiation under current-driven conditions. The responsivity increased mainly caused by the photothermoelectric mechanism; however, the mechanism of the plasmonic detection also exists in a certain wide transcendent range of the applied ADGG bias voltages. Within the applied bias voltage range for the ADGG, the current-driven photothermoelectric detection exhibited superior responsivity almost twice as high as that for the current-driven plasmonic detection while preserving the fast response speed. Moreover, the EG-FET using epitaxial graphene on a SiC substrate is a suitable device for mass productive integrated device process technology. Recently, a novel technique to synthesize high-quality few-layer epitaxial graphene on a single-crystalline SiC thin film grown on a Si wafer has been developed [31]. The precise control of the epitaxially grown graphene layers is also a remaining issue, which could be managed by introducing a microfabricating SiC substrate technique that could spatially confine the epitaxy area [30]. In conclusion, the obtained results indicate that the ADGG-EG-FET THz detector has a promising potential to apply to the 6G- and 7G-class high-speed wireless communication systems.

**Acknowledgements:** The device was fabricated at the Nanoelectronics and Spintronics Laboratory, RIEC, Tohoku University, Japan. This research has been partly carried out at the Fundamental Technology Center, Research Institute of Electrical Communication, Tohoku University. This work was supported by JSPS KAKENHI #16H06361, #18H05331, #20K20349, and #21H01380, Japan. These research results were obtained from the Commissioned Research by NICT #01301, JAPAN.

**Supporting Information:** Derivation of the field-effect mobility of graphene carriers.

**FIGURES**

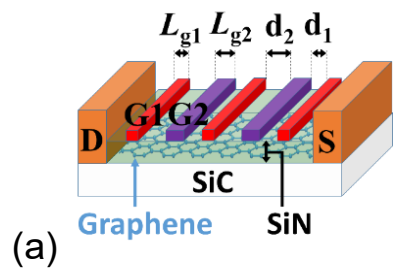

(a)

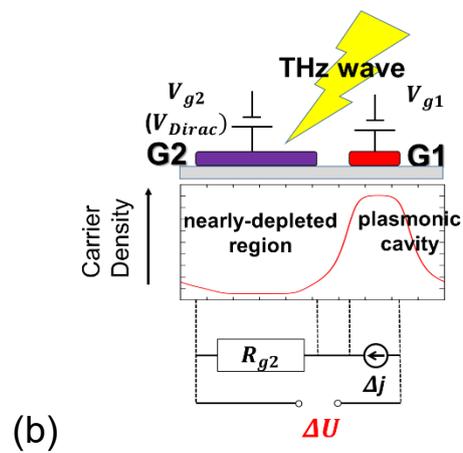

(b)

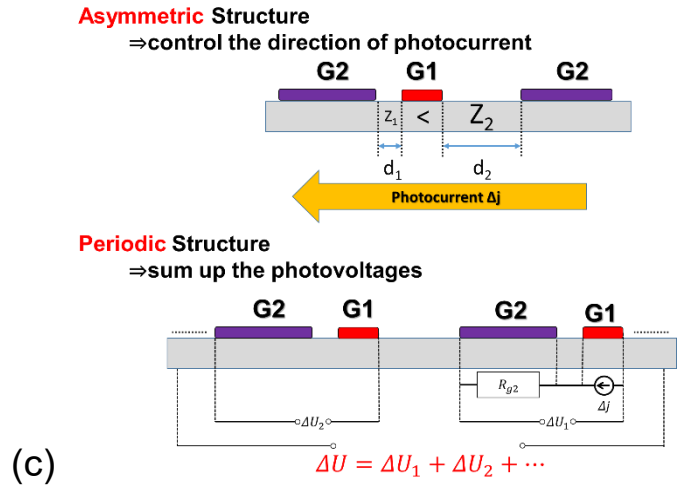

**Figure 1.** Device structure and THz detection mechanisms of an ADGG-EG-FET. (a) A schematic bird's-eye view of the ADGG-EG-FET. The G1 and G2 are biased to appropriate voltages. (b) A schematic image of carrier density distribution in the single ADGG unit in the source-drain direction. A high bias voltage is applied to the electrodes $G_1$ to accumulate the carriers in the channel underneath $G_1$ whereas a low bias voltage is applied to the electrodes $G_2$ to deplete the carriers in the channel underneath $G_2$, forming a periodically arranged high and low carrier density region, in which the plasmonic rectification is realized. When the THz pulse is incident to the EG-FET, the ADGG works as a broadband THz antenna, and the area under G1 (G2) acts as a plasmonic cavity (a resistive load) to generate a plasmonic current (a photovoltage $\Delta U$). (c) The periodically arranged ADGG units. An asymmetric arrangement among the units ($d_1$, $d_2$), ensuring the different impedance on the left and right sides of the plasmonic cavity in the channel underneath the G1 electrode, controls the direction of the photocurrent through the graphene channel. Besides, the periodic structure is able to sum up the photovoltage generated from each unit and obtain the giant output intensity.

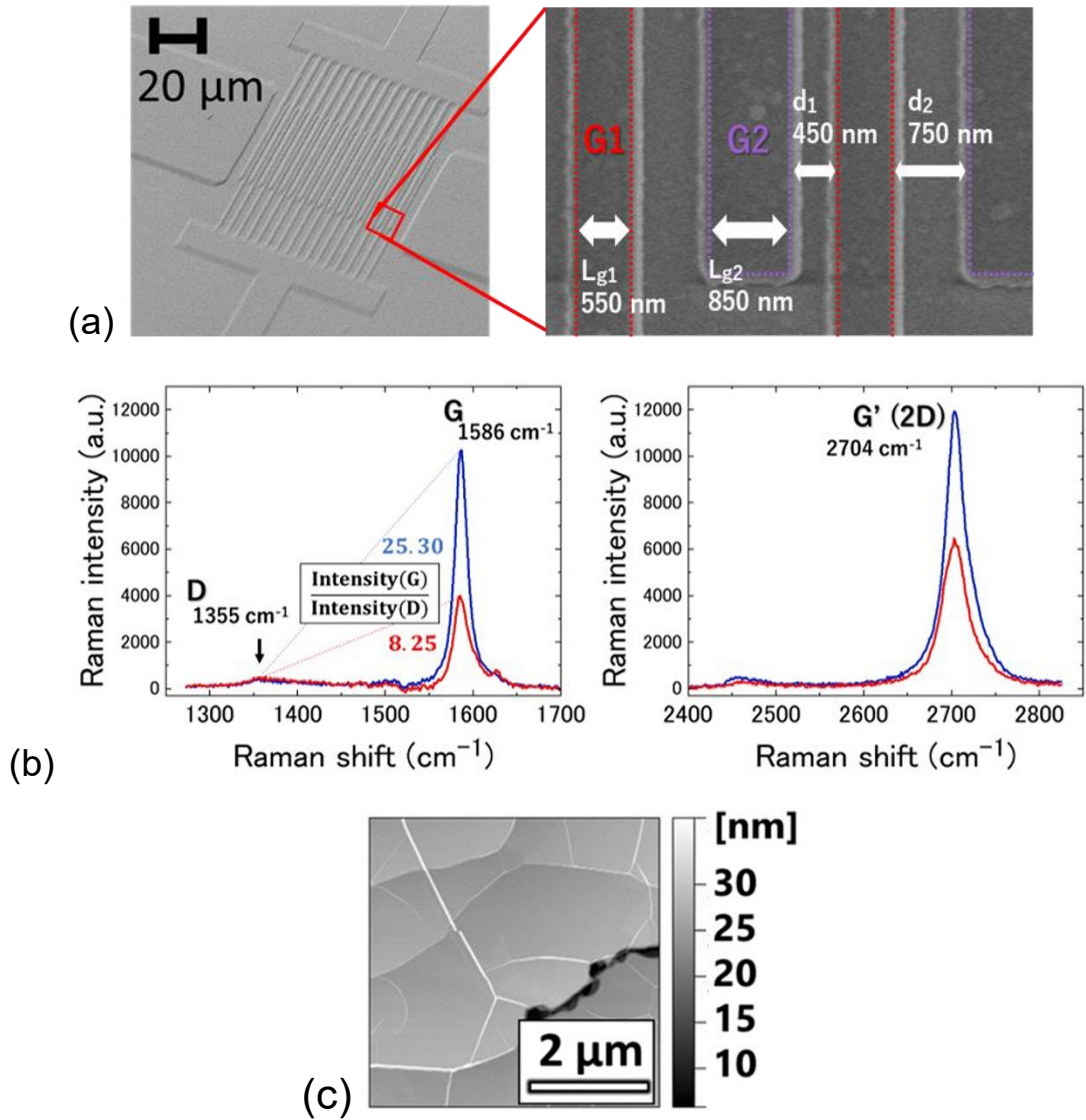

**Figure 2.** The fabricated ADGG-EG-FET THz detector. (a) Scanning electron micrographic images of a fabricated ADGG-EG-FET. The high-magnification SEM image is the observation of an area marked by a red rectangle, by which the size parameters of the device ($L_{g1}$, $L_{g2}$, $d_1$, and $d_2$) and the white-colored SiN gate dielectric layer between the flat epitaxial graphene layer and the gate (G1 and G2) metallization layer can be confirmed. (b) Raman spectra of epitaxial graphene grown on a C-face 6H-SiC substrate, with ranges of from 1300 $cm^{-1}$ to 1700 $cm^{-1}$ (left) and from 2400 $cm^{-1}$ to 2800 $cm^{-1}$ (right) in arbitrary unit. The red curve shows the spectrum with the minimum

ratio of G peak and D peak intensity, and the blue curve shows the spectrum with the maximum ratio. (c) An atomic-force microscopic (AFM) image.

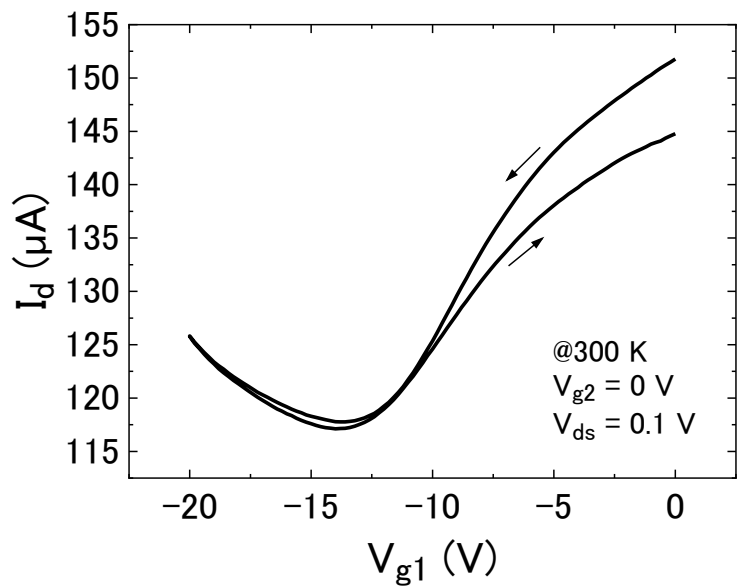

**Figure 3.** Measured ambipolar property of ADGG-EG-FET the gate voltage-dependent dark channel current by double sweep test.

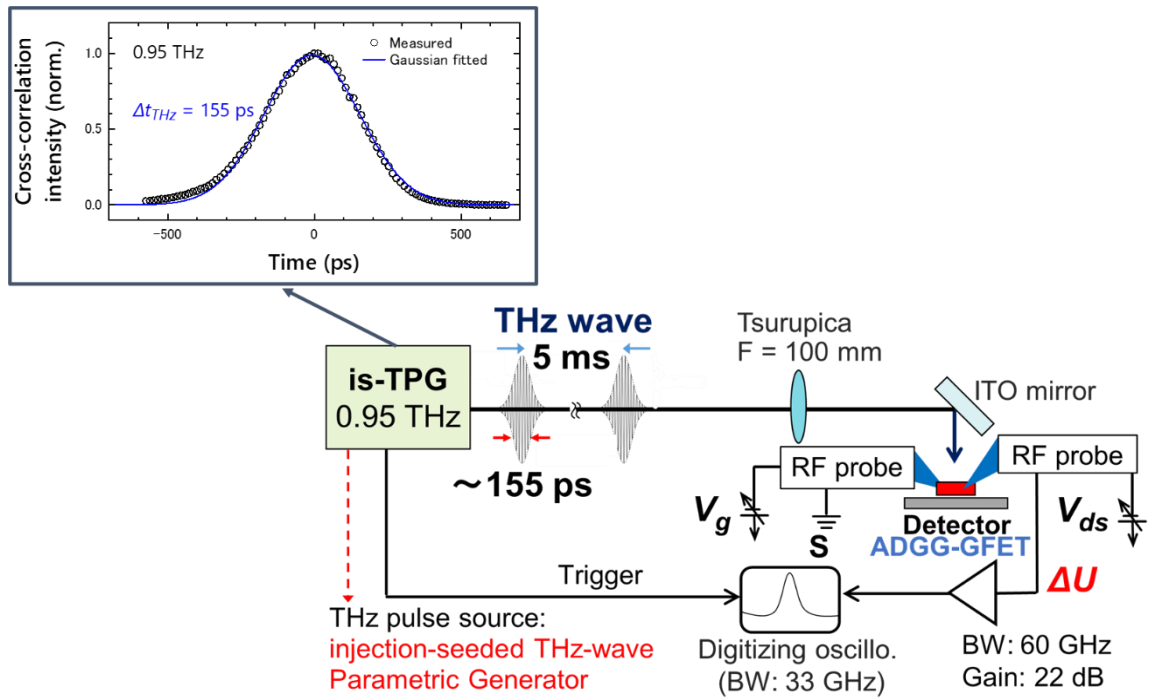

**Figure 4.** Experimental setup for time-domain THz detection measurement. The inset is a typical optically up-converted cross-correlation waveform of the is-TPG THz radiation at 0.95 THz with a sub-ns FWHM Nd:YAG infrared pump pulse centered at 1,064 nm. The FWHM $\Delta t$ of the deconvolved intensity waveform of the is-TPG was identified to be 155 ps.

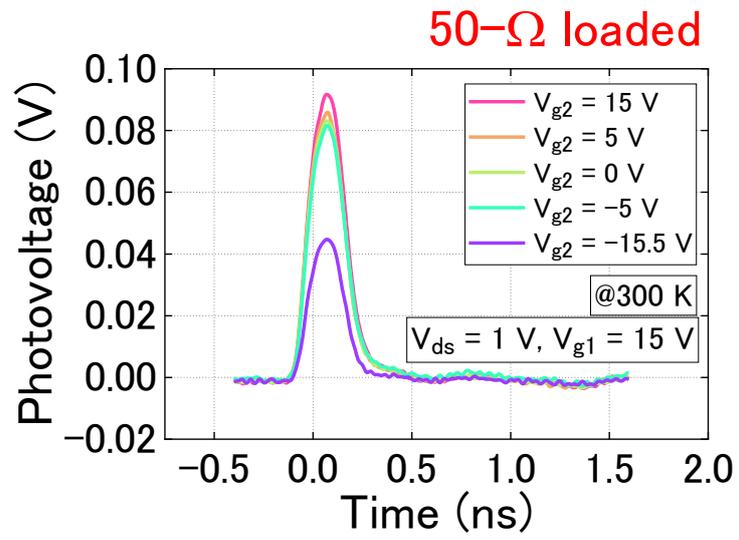

**Figure 5.** Temporal response of output photovoltages for different $V_{g2}$ biases.

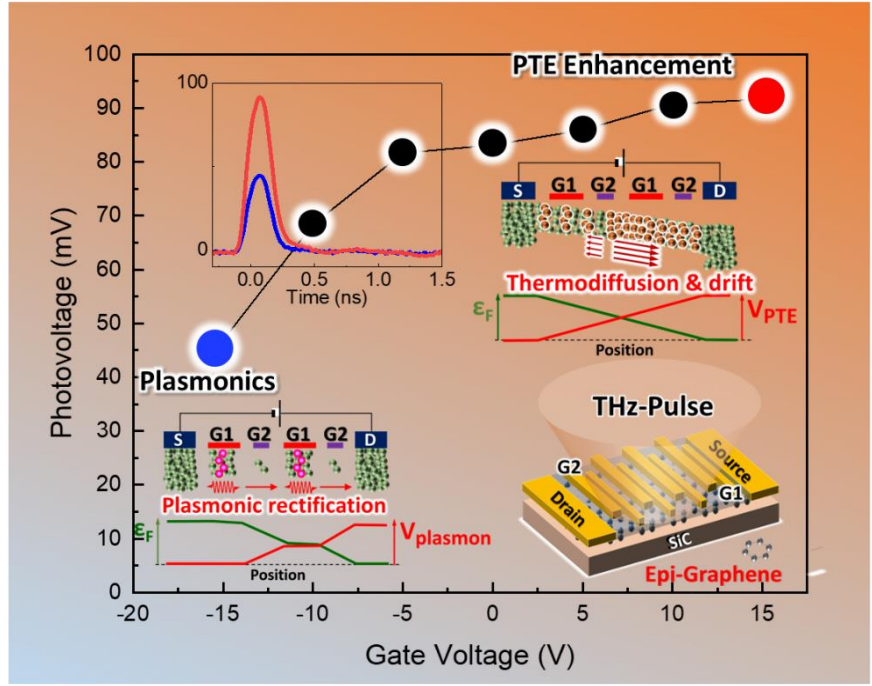

(a)

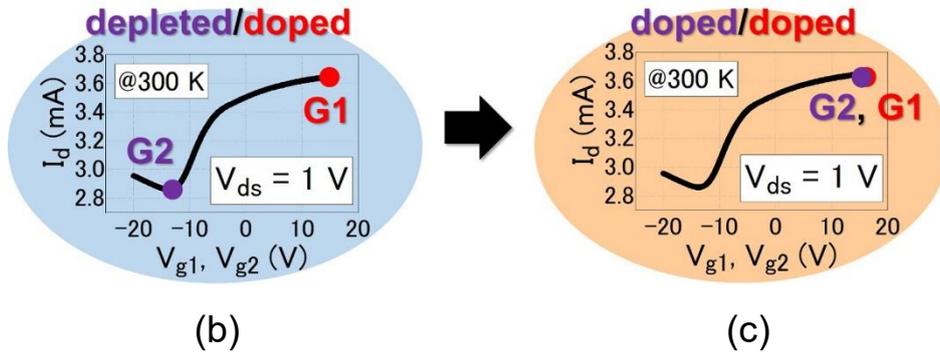

(b)                         (c)

**Figure 6.** The dependence of the THz photoresponse on ADGG biases. (a) The maximum photovoltage versus gate2 bias $V_{g2}$. The schematic views highlight the carrier motions along the channel under the plasmonic and phototermoelectric rectifications. The insert is temporal photoresponse waveforms under the plasmonic (blue line) and phototermoelectric (orange line) rectifications. Two schematics in the blue-shaded oval and orange-shaded oval areas show the plasmonic and photothermoelectric detection mechanisms, respectively. In the plasmonic detection schematic, the pink-colored dots show spatially displaced electrons under THz radiation incidence to promote nonlinear

longitudinal plasma waves of the GDPs that are visualized by the red-colored wavy arrows. The green-colored dots show quasi-equilibrated electrons. In the photothermoelectric detection schematic, the red-colored dots show thermos-diffusive hot electrons due to the photo-Seebeck effect under THz radiation incidence. Application of a dc drain bias voltage gives a potential slope along the channel to make the spatial thermal diffusion to be anisotropic and to weight toward the drain electrode that is visualized by the red-colored arrows. (b) Biasing points for Gate 1 (G1) and Gate 2 (G2) to mediate the plasmonic rectification. (c) Biasing points for G1 and G2 to mediate the photothermoelectric rectification.

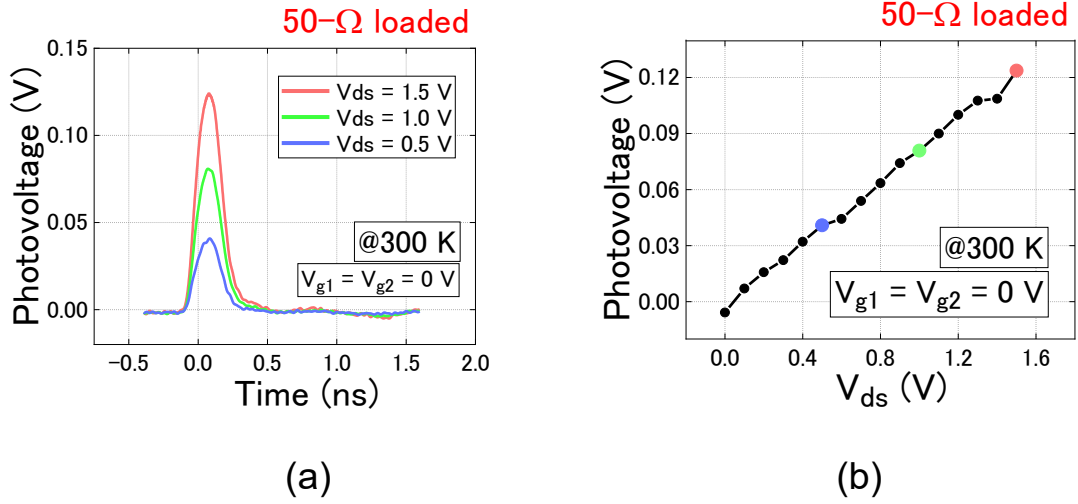

(a)                                        (b)

**Figure 7.** The measured photoresponse under the 50-$\Omega$-loaded condition to the 0.95-THz radiation incidence. (a) Temporal response of output photovoltages with different drain-source biases $V_{ds}$. (b) The maximum photovoltage versus $V_{ds}$.

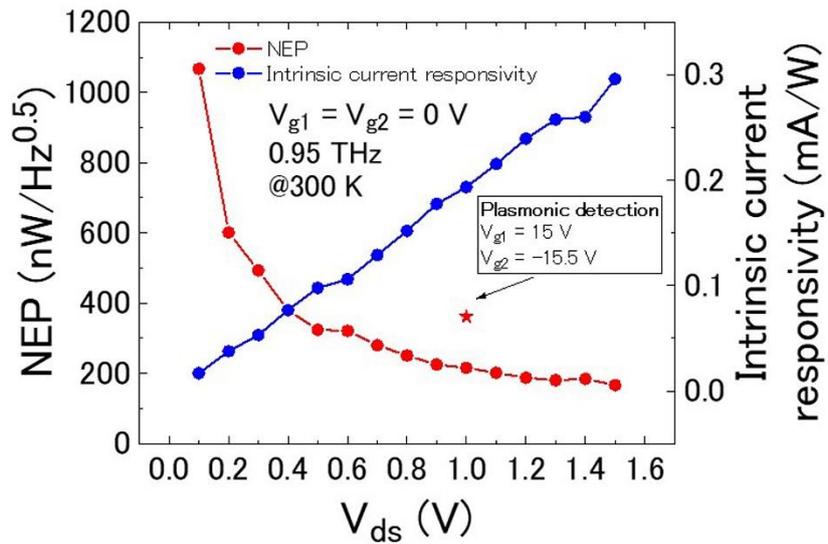

**Figure 8.** *NEP* and the intrinsic current responsivity versus drain-source bias $V_{ds}$.